\documentclass[
    ,final            
  ]
  {aipproc}

\layoutstyle{6x9}

\begin{document}

\title{On the Mass Eigenstate Purity of $^8$Boron 
Solar Neutrinos\footnote{Presented by Stephen Parke at PANIC05, October 
24-28, 2005 in Santa Fe, NM, USA.
}}


\author{Stephen Parke}{
  address={Theoretical Physics Department, Fermi National Accelerator Laboratory,\\
  P.O. Box 500, Batavia, IL 60510, USA\\ email: parke@fnal.gov\\[0.1cm]}
}

\author{Hiroshi Nunokawa}{
  address={\sl Departamento de F\'{\i}sica, Pontif{\'\i}cia Universidade Cat{\'o}lica do Rio de Janeiro, \\
  C. P. 38071, 22452-970, Rio de Janeiro, Brazil\\
  email: nunokawa@fis.puc-rio.br\\[0.1cm]}
}

\author{Renata  Zukanovich Funchal}{
  address={Instituto de F\'{\i}sica, Universidade de S\~ao Paulo,\\C.\ P.\
  66.318, 05315-970 S\~ao Paulo, Brazil\\
  email: zukanov@if.usp.br}
  }
  
\begin{abstract}
 We give a brief report on our recent paper,  Ref. \cite{NPZ2},  in which we calculate the $\nu_2$
mass eigenstate purity of $^8$B solar neutrinos as 91$\pm$2\%.
\end{abstract}

\maketitle



In a two neutrino analysis, the {\it day-time} CC/NC of SNO, which is equivalent to 
the day-time average $\nu_e$ survival probability, $\langle P_{ee} \rangle$, 
is given by
\begin{eqnarray} 
\left. \frac{{\rm CC}}{{\rm NC}}\right|_{{\rm day}} &=&  
\langle P_{ee} \rangle = 
f_1 \cos^2\theta_\odot + f_2 \sin^2 \theta_\odot   
\label{cctonc}, 
\end{eqnarray}
where $f_1$ and $f_2=1-f_1$ are understood to be the $\nu_1$ and $\nu_2$ 
fractions, respectively, averaged over the $^8$B neutrino energy spectrum
weighted with the charged current cross section.
Therefore the $\nu_1$ fraction (or how much $f_2$  differs from 100\%) is given by
\begin{eqnarray} 
f_1 & = & { \left( \left. \frac{{\rm CC}}{{\rm NC}}\right|_{{\rm day}} - \sin^2 \theta_\odot \right)
\over  \cos 2 \theta_\odot } = { (0.347-0.311) \over 0.378 } \approx 10~ \pm ~?? ~\%,
\label{f1eqn}
\end{eqnarray}
where the central values of the recent SNO analysis, \cite{SNO}, have been used.
Due to the correlations in the uncertainties between the CC/NC ratio and $\sin^2\theta_\odot$
we are unable to estimate the uncertainty on $f_1$ from their analysis.
Note, that if the fraction of $\nu_2$ were 100\%,
then $\frac{{\rm CC}}{{\rm NC}}=\sin^2\theta_\odot$.
\begin{figure}[t]
\includegraphics[height=.3\textheight]{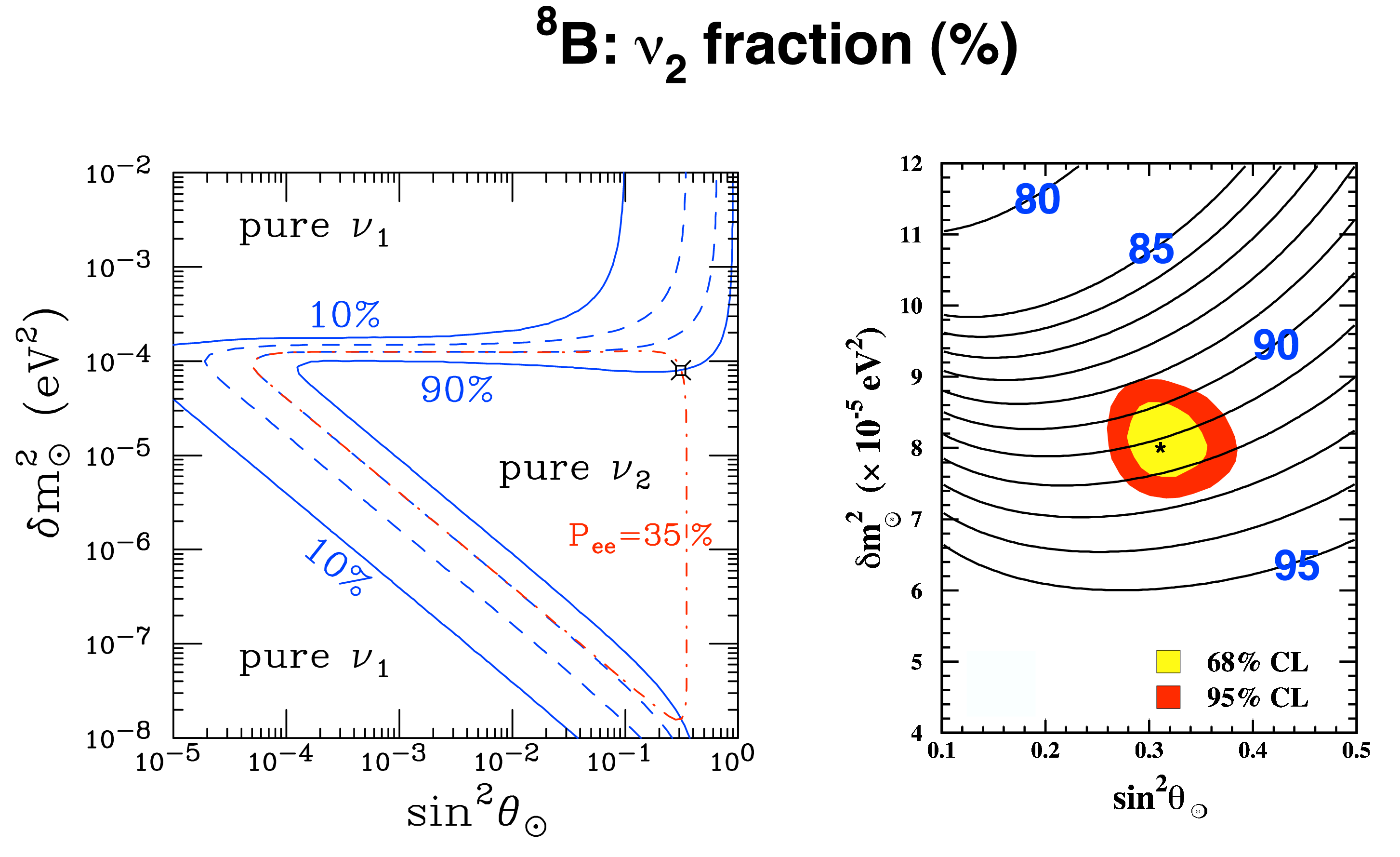}
\caption{The $\nu_2$ fraction (\%) in the $\delta m^2_\odot$ versus $\sin^2
\theta_\odot$ plane. (a) The solid and dashed (blue) lines are the 90, 65, 35 and 10\% 
iso-contours of the fraction of the solar $^8$B neutrinos that are $\nu_2$'s.
  The current best fit value, indicated by the
  open circle with the cross, is close to the 90\% contour. The iso-contour for an electron 
  neutrino survival probability, $\langle P_{ee} \rangle $, equal to 35\% is the dot-dashed (red)
  ``triangle'' formed by the 65\% $\nu_2$ purity contour for small $\sin^2 \theta_\odot$ 
  and a vertical line in the pure $\nu_2$ region at
  $\sin^2 \theta_\odot=0.35$. 
  Except at the top and bottom right hand corners of this triangle 
  the  $\nu_2$ purity is either 65\% or 100\%.
  (b) Focuses in on the current
  allowed region. The 68 and 95\% CL are shown
  by the shaded areas with the best fit values indicated by the star  
  using the combined fit of KamLAND and solar neutrino
  data~  given in \cite{SNO} .   }
\label{nu1nu2}
\end{figure}

Using the analytical analysis of the Mikheyev-Smirnov-Wolfenstein (MSW) effect given in Ref.~\cite{Parke86},
the mass eigenstate fractions are given by
\begin{eqnarray}
f_2 =1 -f_1 & = & \langle  \sin^2 \theta_\odot^N 
+ P_x\cos 2 \theta_\odot^N \rangle_{^8{\rm B}},
\end{eqnarray}
where $\theta_\odot^N$ is the mixing angle defined at the $\nu_e$
production point and $P_x$ is the probability of the neutrino to jump
from one mass eigenstate to the other during the Mikheyev-Smirnov resonance
crossing.  
The average $\langle \cdots \rangle_{^8{\rm B}}$ 
is over the electron density of the
$^8$B $\nu_e$ production region in the center of the Sun
predicted by the Standard Solar Model
and the energy spectrum of $^8$B neutrinos weighted with SNO's charged current cross section.
Fig.~\ref{nu1nu2} shows the iso-contours of this averaged $\nu_2$ fraction using a threshold of 5.5 MeV on the kinetic energy of the recoil electrons.
Thus the $^8$B energy weighted average fraction of $\nu_2$'s observed 
by SNO is 
\begin{equation}
f_2 = 91\pm2\% \quad {\rm at ~the ~95\%~ CL.}
\end{equation}
Hence, the $^8$B solar neutrinos are the purest mass eigenstate neutrino
beam known so far.

\begin{figure}[th]
\includegraphics[height=.45\textheight]{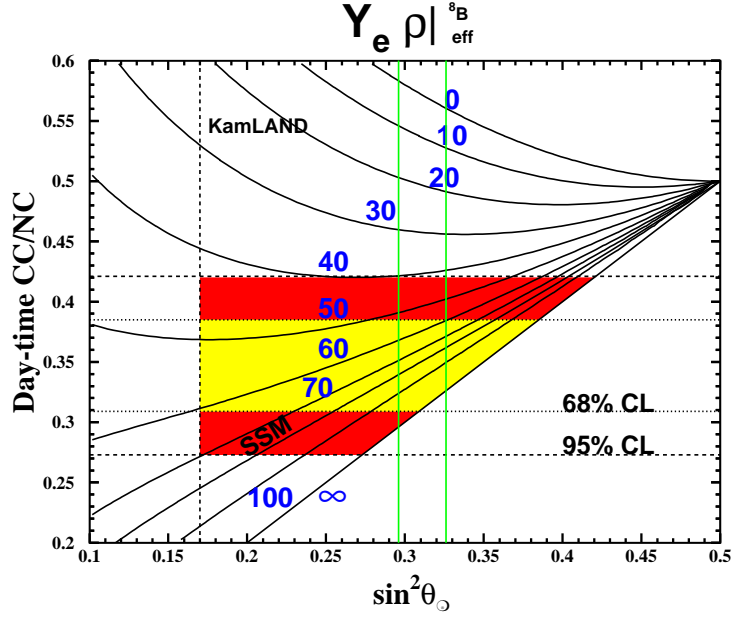}
\caption{The iso-contours of $Y_e\rho~\vert^{^8{\rm B}}_{eff}$
in the $\sin^2\theta_{\odot}-{\rm CC/NC}|_{\rm day}$ plane.
The line labeled SSM is the Standard Solar Model prediction for
$Y_e\rho~\vert^{^8{\rm B}}_{eff}$ ($\approx$ 85 g/cm$^3$).
SNO's range of observed values of CC/NC are indicated by 
the shaded horizontal bands. 
}
\label{yerho_eff}
\end{figure}

Since the propagation of $^8$B neutrinos, in the Sun,
is highly adiabatic, the fraction of $\nu_2$, and
consequently, the SNO CC/NC ratio is determined only by the
effective value of the matter potential at neutrino production near the center of the Sun.
This implies that if we can measure $\sin^2\theta_\odot$  using an experiment independent
of the $^8$B solar neutrinos, 
then from the measured value of SNO's CC/NC ratio
we can determine the effective value of the matter potential at production.
One cannot extract information
on the electron number density distribution or the $^8$B neutrino production
distribution, separately, but only a single characteristic
of the convolution of these two distributions,
$Y_e\rho~\vert^{^8{\rm B}}_{eff}$. Fig. \ref{yerho_eff} demonstrates this observation
and puts a lower bound on  $Y_e\rho~\vert^{^8{\rm B}}_{eff} > 40 g/cm^3$,
see Ref. \cite{NPZ2} for details.

If $\sin^2 \theta_{13}$ is non-zero then we must use a three neutrino analysis.
The three neutrino
mass eigenstate fractions,  $\mathcal{F}_i$ are 
\begin{eqnarray}
\mathcal{F}_1 & = &  f_1   = 0.09\mp 0.02, \nonumber \\
\mathcal{F}_2 & = & f_2 - \sin^2 \theta_{13} 
\approx  0.91\pm 0.02 -\sin^2\theta_{13}, \\
\mathcal{F}_3 & = & \sin^2 \theta_{13}. \nonumber
\end{eqnarray}
A Taylor series expansion about the two flavor mass eigenstate 
fractions, $f_i$, has been used
since $\sin^2 \theta_{13}$ is known to be less than 0.04.

In summary, the $^8$B solar neutrinos are produced and propagate from the center of sun
to the earth's surface as almost a pure $\nu_2$ mass eigenstate with a purity between 85 and 93\%.



\begin{thebibliography}{99}

\bibitem{NPZ2}
H.~Nunokawa, S.~Parke and R.~Z.~Funchal,
``What fraction of B-8 Solar neutrinos arrive at the Earth as a $\nu_2$ mass eigenstate",
Fermilab-Pub-05-049, arXiv:hep-ph/0601198.


\bibitem{SNO}
B.~Aharmim {\it et al.}  [SNO Collaboration],
arXiv:nucl-ex/0502021.


\bibitem{Parke86}
S.~J.~Parke, Phys.\ Rev.\ Lett.\  {\bf 57}, 1275 (1986);
S.~J.~Parke and T.~P.~Walker, Phys.\ Rev.\ Lett.\  {\bf 57}, 2322 (1986).

\end{thebibliography}
\end{document}